\documentclass[pra,floatfix,showpacs,amsmath,twocolumn]{revtex4}
\usepackage{amssymb}
\usepackage{graphics}
\usepackage{epsfig}
\usepackage{graphicx}
\usepackage{psfrag}

\begin{document}

\newcommand{\be}{\begin{equation}}
\newcommand{\ee}{\end{equation}}
\newcommand{\bea}{\begin{eqnarray}}
\newcommand{\eea}{\end{eqnarray}}
\newcommand{\bma}{\begin{subequations}}
\newcommand{\ema}{\end{subequations}}
\def\lR{l^2_{\mathbb{R}}}
\def\RR{\mathbb{R}}
\def\E{\mathbf e}
\def\D{\boldsymbol \delta}
\def\S{{\cal S}}
\def\T{{\cal T}}
\def\dd{\delta}
\def\one{{\bf 1}}
\def\ss{\boldsymbol \sigma}

\newtheorem{theorem}{Theorem}
\newtheorem{lemma}{Lemma}
\newtheorem{definition}{Definition}

\title{Valence Bond Solids for Quantum Computation}

\author{F. Verstraete and J.I. Cirac} \affiliation{Max-Planck-Institut f\"ur Quantenoptik,
Hans-Kopfermann-Str. 1, D-85748 Garching, Germany.}

\date{\today}

\begin{abstract}
Cluster states are entangled multipartite states which enable to
do universal quantum computation with local measurements only. We
show that these states have a very simple interpretation in terms
of valence bond solids, which allows to understand their
entanglement properties in a transparent way. This allows to
bridge the gap between the differences of the measurement-based
proposals for quantum computing, and we will discuss several
features and possible extensions.
\end{abstract}

\maketitle

The concept of teleportation \cite{tele} plays a crucial role in
the understanding of entangled quantum systems. It does not only
allow us to use entangled states as perfect quantum channels, but
also to implement non-local unitary operations
\cite{nielsenchuang}. Based on this idea it was shown that
universal quantum computation can be achieved if one can prepare a
separable initial state and implement joint two-qubit measurements
\cite{gottesman,KLM,niel,leung}. In the same spirit, but somehow
orthogonal to these schemes, Raussendorf and Briegel
\cite{Raussendorf} showed that universal quantum computation is
possible by doing local measurements on the qubits in a highly
entangled so--called cluster state \cite{Briegel}. These studies
highlighted the central role of entanglement for quantum
computation \cite{Jozsa,Vidal}. However, the structure of general
multiparticle entanglement is, for the moment being, still very
poorly understood, and it is somehow mysterious that the cluster
states enable for universal quantum computation. In this note, we
show that the structure of entanglement in cluster states is
particularly simple and can be well understood by looking at it as
a so--called valence bond solid with only nearest-neighbor bonds.
This enables us to show that the one-way computer
\cite{Raussendorf} essentially works in an equivalent way as the
other measurement-based proposals for quantum computation.

\begin{figure}[t]
\includegraphics[width=8cm]{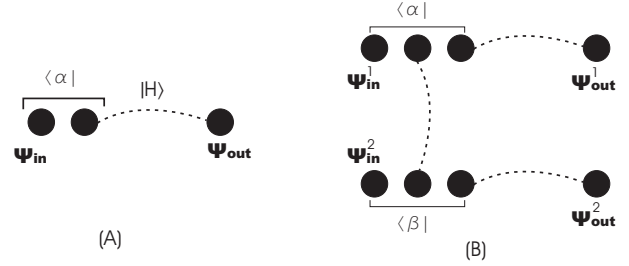}
\caption{(A) Implementation of a 1-qubit gate by measuring in the
2-qubit basis $|\alpha\rangle$ (\ref{meas1}). The edges connected
by the line denote the maximally entangled state $|H\rangle$. (B)
Implementation of a 2-qubit gate by 3-qubit measurements in the
basis $|\alpha\rangle$ and $|\beta\rangle$
(\ref{meas2}).}\label{Uph}
\end{figure}

Let us start by recalling how universal quantum computation can be
performed using measurements only on a collection of maximally
entangled states of 2 qubits. Schematically, we will present the
logical qubits as the leftmost particles. Logical gates can be
implemented by introducing extra pairs of maximally entangled
states $|H\rangle=|00\rangle+|01\rangle+|10\rangle-|11\rangle$ of
two qubits (every other maximally entangled would also be good),
and doing Bell measurements on halves of these $|H\rangle$ and the
logical qubits. The new logical qubits are now in the other parts
of the maximally entangled state. It is well known that a
universal set of gates \footnote{Note that universal quantum
computation can be achieved by only implementing gates between
neighboring qubits.} is given by arbitrary local unitary
transformations and the phase gate $U_{ph}=|00\rangle\langle
00|+|01\rangle\langle 01|+|10\rangle\langle 10|-|11\rangle\langle
11|$. A local unitary transformation $U$ on qubit $i$ can be
implemented by doing a Bell measurement in the basis
\be\label{meas1}|\alpha\rangle=(U^\dagger\sigma_\alpha\otimes
\openone)|H\rangle,\hspace{.2cm}\alpha=0,1,2,3\ee where
$\sigma_\alpha$ denote the Pauli matrices (including
$\sigma_0=\openone$); see Figure \ref{Uph}A. This implements the
unitary gate $\sigma_\alpha U$ which is $U$ up to an extra
multiplication with a Pauli operator ($\alpha$ is conditioned by
the measurement outcome); this extra Pauli operator however does
not harm \footnote{The extra left multiplication with Pauli
operators does not harm: a later 1-qubit operation $V$ can be
chosen to be conditioned on the outcome (i.e. implementing
$V\sigma_\alpha$ instead of $V$). Furthermore, right
multiplication of the 2-qubit phase gate $U_{ph}$ with Pauli
operators is equivalent to left multiplication of it with
different ones. Therefore the extra Pauli operators can be pushed
through the quantum circuit without affecting the computation.}.
Similarly, the phase gate $U_{ph}$ can be implemented by adding
three extra pairs of maximally entangled states $|H\rangle$ as
depicted in Figure \ref{Uph}B. Suppose two three-qubit
measurements are done (see Fig. \ref{Uph}B) in the complete bases
\be\label{meas2}\{|\alpha\rangle\}=\{|\beta\rangle\}=(\sigma_x)^i\otimes
(\sigma_x)^j\otimes
\openone(|0\rangle|0\rangle|0\rangle\pm|1\rangle|1\rangle|1\rangle)\ee
with $i,j\in\{0,1\}$ and
$|\pm\rangle=(|0\rangle\pm|1\rangle)/\sqrt{2}$. It can readily be
checked that this implements the gate $(H\otimes H)U_{ph}$  with
$H$ the Hadamard gate $H=|+\rangle\langle 0|+|-\rangle\langle 1|$,
up to a harmless extra multiplication with Pauli operators.
Together with the possibility of implementing local unitaries,
this proves that universal quantum computation can be done by
doing only Bell measurements on two or three qubits.

Let us summarize the three ingredients needed for being able to
implement quantum computing along the lines sketched: 1/ it must
be possible to create ancillary singlets; 2/ two- and three-qubit
measurements of the form (\ref{meas1},\ref{meas2}) can be
implemented between halves of these extra singlets and the logical
qubits.

\begin{figure}[t]
\includegraphics[width=8cm]{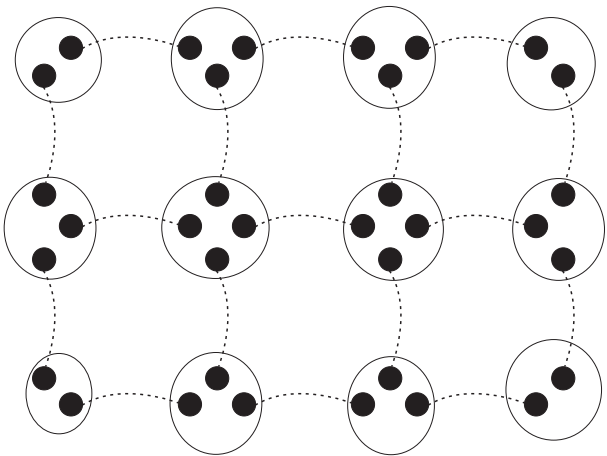}
\caption{Representation of a Valence Bond Solid. The edges
connected with a dotted line denote a singlet, while the circles
denote a projection $P$ of all qubits inside it with Hilbert space
$H_2^{\otimes n}$ to a single qubit $\tilde{H}_2$. In the present
paper, $P$ is always of the form $P=|\tilde{0}\rangle\langle
00\ldots 0|+|\tilde{1}\rangle\langle 11\ldots 1|$.}\label{VBS}
\end{figure}

Inspired by the AKLT-valence bond solids (VBS)
\cite{AKLT,fannes,Michelangelo}, playing a central role in
condensed matter physics, it is now  interesting to investigate
whether it is possible to interpret the two (or three) qubits on
which the measurements have to be implemented as virtual qubits
representing one physical qubit. As will become clear, this will
exactly lead to the concept of the one-way computer
\cite{Raussendorf}. As depicted in Figure \ref{VBS}, VBS are
constructed by distributing singlets $|H\rangle$ made from virtual
qubits between different sites, followed by a local projection of
the virtual qubits on a smaller dimensional subspace $\tilde{H}$
encoding the physical qubit. In the case of the 2-D lattice in
Figure \ref{VBS} e.g., the 4 qubits can be projected on a one
qubit subspace by the operator $P_4=|\tilde{0}\rangle\langle
0000|+|\tilde{1}\rangle\langle 1111|$ (here 4 specifies that there
are 4 virtual qubits; $P_n$ is defined as
$P_n=|\tilde{0}\rangle\langle 00\ldots 0|+|\tilde{1}\rangle\langle
11\ldots 1|$ with $n$ virtual qubits). The idea is thus to
interpret one of the virtual qubits as the logical one, and the
other virtual ones as the ones enabling teleportation.

Let us see whether the two listed requirements for measurement
based quantum computation can be fulfilled.

1/ Consider the VBS of Figure \ref{VBS}. To implement a virtual
2-qubit (3-qubit) measurement, we want that the physical qubit is
only made up of 2 (3) virtual qubits (on which we want to
implement a Bell measurement), and hence that 2 (1) virtual
qubit(s) effectively disappear. Suppose the physical qubit
(Hilbert space $\tilde{H}$) is obtained by projecting the virtual
ones by the projector $P_4=|\tilde{0}\rangle\langle
0000|+|\tilde{1}\rangle\langle 1111|$. Measuring 2 (1) neighboring
physical qubit(s) in the
$|\tilde{0}\rangle,|\tilde{1}\rangle$-basis effectively replaces 2
(1) virtual qubit(s) with $|+\rangle$ or $|-\rangle$ depending on
the measurement outcome. But it holds that
$P_4|+\rangle=P_3\equiv|\tilde{0}\rangle\langle
000|+|\tilde{1}\rangle\langle 111|$ and
$P_3|-\rangle=\tilde{\sigma}_zP_3$, and similarly
$P_3|+\rangle=P_2$. Therefore the bonds in a VBS can be broken at
will by measuring neighboring physical qubits, the virtual qubits
effectively disappear and the projector $P_n$ changes into
$P_{n-1}$.

2/ Let us now investigate whether virtual measurements such as
(\ref{meas1},{meas2}) can be implemented by local measurements on
physical qubits. This will be possible if the projector $P_n$ has
full support on a subset of rays corresponding to these multiqubit
measurements. Clearly, this is the case for the phase gate if the
projector is $P_3$, as a measurement in the physical basis
$|\tilde{0}\rangle\pm|\tilde{1}\rangle$ effectively corresponds to
the measurement of the virtual qubits in the basis
$|000\rangle\pm|111\rangle$ which belongs to the optimal basis
(\ref{meas2}). In the case of the single qubit gates, things are a
bit more subtle. Suppose the projector is $P_2$. Only measurements
in bases of the form
$|\tilde{0}\rangle\pm\exp(i\xi)|\tilde{1}\rangle$ correspond to
Bell measurements, and its effect is to implement the unitary
\[U=(\sigma_x)^{k}\frac{1}{\sqrt{2}}\left(%
\begin{array}{cc}
  1 & \exp(-i\xi) \\
  1 & -\exp(-i\xi) \\
\end{array}%
\right)\] on the (virtual) logical qubit ($k=0,1$ depending on the
measurement outcome). It can however easily be checked that a
sequence of 4 such unitaries can generate every specified unitary
$\in SU(2)$ \footnote{See e.g. the way 1-qubit unitary gates are
implemented in the one-way computer \cite{Raussendorf}.} (this
again holds up to left multiplication with Pauli matrices).
Therefore the second requirement is also fulfilled.

By measuring the qubits from left to right on a lattice, the
(virtual) logical qubits travel from left to right, yielding
quantum computation using single-qubit measurements only. This
shows that 2-D valence bond solids can be used to do a quantum
computation. This model of computation exactly coincides with the
one-way computer of Rausendorf and Briegel; indeed, we will next
show that the cluster state is exactly the VBS used in the above
construction.

\begin{figure}[t]
\includegraphics[width=5cm]{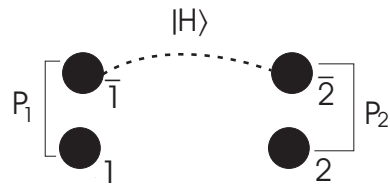}
\caption{Implementing a global unitary
transformation on qubits 1 and 2 by doing local projections $P_1$
and $P_2$ on them and a maximally entangled state
$|H\rangle$.}\label{tele}
\end{figure}

The cluster states are a subset of the so--called stabilizer
states \cite{gott96}, which are defined by specifying a complete
set of commuting observables $O_i$, where each $O_i$ is a tensor
product of the Pauli matrices
$\sigma^0=\openone_2,\sigma^x,\sigma^y,\sigma^z$. The stabilizer
states are the common eigenstates of these operators. Let us show
that any stabilizer state can be interpreted as a valence bond
state. Stabilizer states can efficiently be prepared from a
completely separable state by applying appropriate 2-qubit unitary
operations to it (see e.g. \cite{NC}). The reason that stabilizer
states are very simple and manageable to work with is due to the
fact all these 2-qubit unitary operations can be chosen to commute
with each other. This can readily be seen by looking at the normal
form for stabilizer states \cite{gotthesis}, and then making use
of properties of Pauli operators such as
$U(\sigma^x\otimes\sigma^z)U^\dagger=\sigma^x\otimes I$ with
$U={\rm diag}[1,1,1,-1]$ to diagonalize all operators $O_i$. The
trick is now to implement these commuting 2-qubit unitary
transformations by a teleportation-like principle that consists of
adding virtual singlets, and then doing appropriate projections
\cite{nielsenchuang,cirac}. More specifically, consider the two
qubits $1$ and $2$ in Figure \ref{tele}; an extra singlet
$|H\rangle_{\bar{1}\bar{2}}$ is added, and then any unitary
transformation between $1$ and $2$ can be simulated by projecting
the two-qubit spaces labelled by $1,\bar{1}$ ($2,\bar{2}$) onto
the qubits $1$ ($2$ with appropriate projectors $P_1$ ($P_2$
($P_1,P_2$ are $2\times 4$ matrices). Iterating this scheme, one
readily sees that every stabilizer state can be interpreted as a
VBS, possibly with bonds extending over all sites \footnote{It
would be interesting in this respect to find a normal form for
stabilizer states that minimizes this number of bonds. This
however seems to be a difficult problem \cite{nor}.}. In the case
of the cluster states however, only unitaries between the nearest
neighbors have to be implemented, and hence a simple VBS as
depicted in Figure \ref{VBS} is obtained.

As an example, let us explicitly construct the valence bond states
corresponding to arbitrary cluster states \footnote{The
construction given also works for the set of so--called graph
states \cite{Schlingemann}, which is, up to local unitaries,
equivalent to the set of stabilizer states.}. To each cluster
state, one can associate a graph parameterized by its adjacency
matrix $\Gamma$. The number of qubits on each site in the VBS is
of course equal to the number of bonds of the given site, and is
equal to the number of vertices emanating from a given physical
qubit. The bonds are maximally entangled states
$|H\rangle=|00\rangle+|01\rangle+|10\rangle-|11\rangle$, and the
projectors on each site are all of the form
$P=|\tilde{0}\rangle\langle 00\ldots 0|+|\tilde{1}\rangle\langle
11\ldots 1|$. This simple construction describes all possible
cluster states.

This VBS interpretation of cluster states makes their nice and
appealing  properties very explicit. The fact that e.g. a singlet
can be created between two arbitrary qubits by doing appropriate
local measurement on the other ones can readily be understood by
the concept of entanglement swapping \cite{swapping}. The entropy
of a block of spins can readily be seen to be given by the number
of bonds emanating from it (i.e. proportional to the area of the
surface of the block). The fact that the sensitivity to noise of a
cluster state does not scale with the number of (physical) qubits
\cite{dur1}, is of due to the fact that it is effectively made up
by \emph{local} singlet pairs. This insight also enables to
construct distillation protocols for cluster states by translating
bipartite distillation protocols to the valence bond picture
\cite{dur2}.

On the other hand, the description of valence bond states in terms
of stabilizer states is also interesting from the point a view of
condensed matter theory. It is e.g. well-known that operations of
the Clifford group acting on a stabilizer state can easily be
simulated efficiently classically. This implies that evolutions
generated by the Clifford group on VBS-states can be simulated
efficiently, and correlation functions of products of Pauli
operators can easily calculated. On the other hand, the
possibility to do quantum computation with VBS using local
measurements only proves that the complexity-class for calculating
general expectation values on 2-D VBS is the same as the
complexity class of quantum computating.

The present study also  opens the question whether there exist
ground states of (gapped) Hamiltonians involving only 2-body
short-range interactions on a lattice that would enable to
implement the presented measurement scheme (this is not the case
for cluster states). Such 2-D Valence Bond Solids indeed exist for
higher spins (e.g. spin $3/2$), and it is trivial do devise a toy
model for which this holds. Consider e.g. a hexagonal lattice with
spin-$7/2$ particles at each vertex. To each particle corresponds
a $8$-dimensional Hilbert space, which we can interpret as a
system of $3$ qubits. We associate each outgoing edge to one of
these qubits, and associate the Hamiltonian
$\vec{S}\vec{S}+3\openone$ to two of these qubits connected by an
edge. One readily sees that the ground state on such a hexagonal
lattice with this 2-body local Hamiltonian will be unique, and
that the teleportation scheme can be implemented perfectly on it.
Note that the cluster state is very similar to that construction,
but there the $3$ qubits are interpreted as virtual qubits and a
smart projection was used to reduce the dimension of the effective
Hilbert space.

More interestingly, the trick used to implement 2-qubit unitary
gates by introducing a virtual singlet followed by a projection -
this is the way cluster states can be generated from completely
separable ones -   can also be extended to the case where the
unitaries do not commute with each other. Indeed, the cluster
state can be made in the lab if an Ising interaction can be
implemented on neighboring qubits \cite{Bloch}. However, in some
experimental setups, it is not always possible to implement such
commuting gates, as is the case e.g. for quantum dots \cite{loss}:
here one is essentially restricted to implement 2-qubit gates
generated by the Heisenberg interaction, which certainly do not
commute when acting on neighboring spins. However, if one can
apply these unitary gates sequentially (i.e. one has control over
the sites on which one implements the gate), then it is also
possible to construct valence bond solids that are suitable for
quantum computation.

The present results also prove that the valence bond solid picture
is very useful for understanding multipartite entanglement.
Indeed, VBS are particularly interesting from the point a view of
quantum information theory, as the simple and elegant tools
developed for bipartite quantum systems can be applied to it (see
e.g. \cite{Michelangelo}). Moreover, one can readily see that the
VBS form a dense subset of all possible quantum states if one
allows the bonds to extend beyond nearest neighbors, if the
singlets are replaced by higher dimensional maximally entangled
states and if the projectors can be chosen arbitrary (e.g. in the
case of 3 qubits, every state can be made by considering 2
singlets and projecting 2 qubits of them onto a qubit space
\cite{Miyake}). It would be very interesting to develop a general
theory of multiparticle entanglement based on this VBS-picture,
where one could construct entanglement measures that quantify the
valence bond resources needed to describe the state. This will be
reported elsewhere.

In conclusion, we have identified the entanglement properties of
the cluster states that are responsible for the possibility of
universal quantum computation. The main insight was given by the
fact that the structure of entanglement in these states is
essentially bipartite and can be understood in terms of valence
bonds. This allowed to prove the equivalence of the one-way
computer with teleportation-based computation schemes, and to
clarify the special features of the cluster states.

\acknowledgements We acknowledge M. Mart\'in--Delgado for
interesting discussions about VBS.


\begin{thebibliography}{99}

\bibitem{tele} C.H. Bennett et al., Phys. Rev. Lett. {\bf 70}, 1895 (1993).

\bibitem{nielsenchuang} M. A. Nielsen and I. L. Chuang, Phys. Rev. Lett. {\bf 79}, 321
(1997).

\bibitem{gottesman} D. Gottesman and I. Chuang, Nature {\bf 402},
390 (1999).

\bibitem{KLM} E. Knill, R. Laflamme and G. Milburn, Nature {\bf
409}, 26 (2001).

\bibitem{niel} M.A. Nielsen, quant-ph/0108020.

\bibitem{leung} D. Leung, quant-ph/0111122 and quant-ph/0310189.

\bibitem{Raussendorf} R. Raussendorf and H.J. Briegel, Phys. Rev.
Lett. {\bf 86}, 5188 (2001); Quant. Inf. Comp. 6, 443 (2002).

\bibitem{Briegel} H. Briegel and R. Raussendorf, Phys. Rev. Lett.
{\bf 86}, 910 (2001).

\bibitem{Jozsa}  R. Jozsa and N. Linden, quant-ph/0201143.

\bibitem{Vidal}  G. Vidal, Phys. Rev. Lett. {\bf 91}, 147902
(2003).

\bibitem{AKLT} I. Affleck, T. Kennedy, E.H. Lieb and H. Tasaki,
Commun. Math. Phys. 115, 477 (1988); Phys. Rev. Lett. 59, 799-802
(1987)


\bibitem{fannes} M. Fannes, B. Nachtergaele and R.F. Werner, Comm. Math. Phys.
{\bf 144}, 443 (1992).

\bibitem{Michelangelo} F. Verstraete, M.A. Mart\'in--Delgado and J.I.
Cirac, quant-ph/0311087.


\bibitem{gott96} D. Gottesman,  Phys.Rev. A {\bf 54}, 1862 (1996).

\bibitem{NC} M. Nielsen and I. Chuang, Quantum computation and quantum
information, Cambridge University Press, Cambridge, U.K (2000).

\bibitem{gotthesis} D. Gottesman, Caltech Ph.D. Thesis,
quant-ph/9705052.

\bibitem{cirac} J.I. Cirac, W. D\"ur, B. Kraus and M. Lewenstein, Phys. Rev. Lett.
{\bf 86}, 544 (2001).




\bibitem{Schlingemann} D. Schlingemann and  R.F. Werner,
Phys. Rev. A {\bf 65}, 012308 (2002).

\bibitem{nor} M. Hein, J. Eisert and H.J. Briegel,
quant-ph/0307130;  M. Van den Nest, J. Dehaene and B. De Moor,
quant-ph/0308151;  J. Dehaene and B. De Moor, Phys. Rev. A {\bf
68}, 042318 (2003).

\bibitem{swapping} M. Zukowski et al., Phys. Rev. Lett. {\bf 71},
4287 (1993).

\bibitem{dur1} W. D\"ur and H.-J. Briegel, quant-ph/0307180.

\bibitem{dur2} W. D\"ur, H. Aschauer and H.-J. Briegel, Phys. Rev. Lett. {\bf 91}, 107903
(2003).

\bibitem{Bloch} O. Mandel et al., Nature {\bf 425}, 937  (2003).

\bibitem{loss} D. Loss and D.P. DiVincenzo, Phys. Rev. A {\bf 57},
120 (1998).



\bibitem{Miyake} A. Miyake and F. Verstraete, quant-ph/0307067.


\end{thebibliography}
\end{document}